# Collective excitations in biased bilayer graphene: Temperature effects

Nguyen Van Men [1,2], Nguyen Quoc Khanh [2,3*], and Dong Thi Kim Phuong [1,2]

[1] University of An Giang - VNUHCM, 18-Ung Van Khiem Street, Long Xuyen, An Giang, Viet Nam (Email: nvmen@agu.edu.vn).
[2] Vietnam National University, Ho Chi Minh City, Viet Nam.
[3] University of Science - VNUHCM, 227-Nguyen Van Cu Street, 5th District, Ho Chi Minh City, Viet Nam (Email: nqkhanh@hcmus.edu.vn).

**Abstract**

We have studied the temperature effect on collective excitations in biased bilayer graphene within random-phase approximation. From the zeros of temperature dynamical dielectric function of the system we have found one weakly damped plasmon mode. For a given electrostatic potential bias, at low (high) temperature $T$ the plasmon frequency changes slightly (increases remarkably) with $T$. We have also studied the effects of potential bias and carrier density on the plasmon frequency of the system at finite temperatures.

## 1. Introduction

Since graphene was experimentally discovered, it has been paid a large amount of attention from scientists due to its potential applications [1-13]. This material consists of only one layer of carbon atoms arranged in a honey-comb lattice, forming an ideal two-dimensional system, called monolayer graphene (MLG). The quasi-particles in MLG are massless chiral Dirac fermions with linear low-energy dispersions in momentum space. Meanwhile, quasi-particles in bilayer graphene (BLG), consisting of two parallel sheets of MLG separated by a small distance, behave as massive fermions with parabolic low-energy dispersions [14-17]. It is proven theoretically and experimentally that an energy gap between conduction and valence band can be generated by an electrostatic potential bias $U$ between two graphene layers. The presence of potential bias results in significant changes of BLG properties such as ground state energy, screening, collective excitations etc. [18-26].

The plasmon excitations in MLG, monolayer gapped graphene, BLG have been investigated intensively with interesting results and many exciting applications have been suggested [14-18, 27-35]. In these studies the authors have either ignored the effects of energy gap or have taken into account the latter caused by the symmetry breaking of A and B sub-lattice, spin-orbit interactions and external magnetic field [35-40]. Recently, the authors of Ref. 18 have calculated the zero-temperature plasmon spectrum in a BLG under a perpendicular electric bias. They have shown that a gap opened by the potential bias can lead to undamped collective excitation modes observable in experiments. Having in mind that the temperature effect is very important in realistic experiments we investigate in this paper the plasmon dispersion relations in a biased bilayer graphene (BBLG) system at finite temperature.

## 2. Theory

The collective excitations in a structure can be found from the zeroes of its dynamic dielectric function [27-34],

$$\varepsilon(q, \omega_p - i\gamma, T) = 0 \qquad (1)$$

where $\omega_p$ is the plasmon frequency at a wave vector $q$ and $\gamma$ is the damping rate of plasma oscillations.

In case of weak damping ($\gamma << \omega_p$), the plasmon dispersion can be determined from the following equation [27-34]

---

* Corresponding author: nqkhanh@hcmus.edu.vn (N.Q. Khanh).

$$\text{Re}\,\varepsilon(q,\omega_p,T) = 0 \tag{2}$$

The temperature dynamic dielectric function of BBLG is written as

$$\varepsilon(q,\omega,T) = 1 - v(q)\Pi_0(q,\omega,T) \tag{3}$$

where $v(q)$ is Coulomb bare interaction in momentum space [27-30, 34]

$$v(q) = \frac{2\pi e^2}{\kappa q} \tag{4}$$

with $\kappa$ being the average dielectric constant of surrounding environment and $\Pi_0(q,\omega,T)$ is the temperature polarization function of BBLG. Within the random-phase-approximation, this function has the form [18]

$$\Pi(q,\omega,T) = g \sum_{\lambda,\lambda',\vec{k}} \left|g_{\vec{k}}^{\lambda,\lambda'}(\vec{q})\right|^2 \frac{f(E_{\vec{k}+\vec{q}}^{\lambda'}) - f(E_{\vec{k}}^{\lambda})}{\omega + E_{\vec{k}+\vec{q}}^{\lambda'} - E_{\vec{k}}^{\lambda} + i\delta}. \tag{5}$$

where the factor $g = 4$ describes the spin and valley degeneracy of the electronic states, and $\left|g_{\vec{k}}^{\lambda,\lambda'}(\vec{q})\right|^2$ is the vertex factor [18],

$$\left|g_{\vec{k}}^{\lambda,\lambda'}(\vec{q})\right|^2 = \frac{1}{2}\left[1 + \lambda\lambda' \cos\alpha_{\vec{k}} \cos\alpha_{\vec{k}+\vec{q}} + \lambda\lambda' \sin\alpha_{\vec{k}} \sin\alpha_{\vec{k}+\vec{q}} \cos(2\theta_{\vec{k}} - 2\theta_{\vec{k}+\vec{q}})\right] \tag{6}$$

with

$$\tan\alpha_{\vec{k}} = \frac{\hbar^2 k^2}{m^* U}. \tag{7}$$

Here $f(x)$ is the Fermi-Dirac distribution function and

$$E_{\vec{k}}^{\lambda} = \lambda \sqrt{\left(\frac{U}{2}\right)^2 + \left(\frac{\hbar^2 k^2}{2m^*}\right)^2} \tag{8}$$

is the energy of electrons in BBLG near Dirac points $K$ and $K'$.

We have solved Eq. (2) numerically using equations (3)-(8) and the obtained results for the finite-temperature plasmon dispersion are given in the next section.

## 3. Results and discussions

In the followings, the Fermi energy, the Fermi wave vector and the Fermi temperature of the unbiased BLG system are denoted by $E_F$, $k_F$ and $T_F$, respectively. Assuming that surrounding media of our system is SiO$_2$ and vacuum we use the average dielectric constant $\kappa = \dfrac{\kappa_{SiO_2} + 1}{2} = \dfrac{3.8 + 1}{2} = 2.4$.

In order to understand the effects of temperature on plasmon properties of the system, we plot in Fig. 1 plasmon frequencies in a BBLG with $n = 10^{12}\,cm^{-2}$, $U = 0.5E_F$ for $T = 0$ and $0.2T_F$ (a), and $T = 0.2$ and $0.4T_F$ (b). It can be seen from the figure that for small (large) wave vectors, the temperature modifies slightly (increases significantly) the plasmon frequency. The temperature effect is more pronounced at higher temperature. Note that our results for $T = 0$ are similar to those given before in Ref. 18.

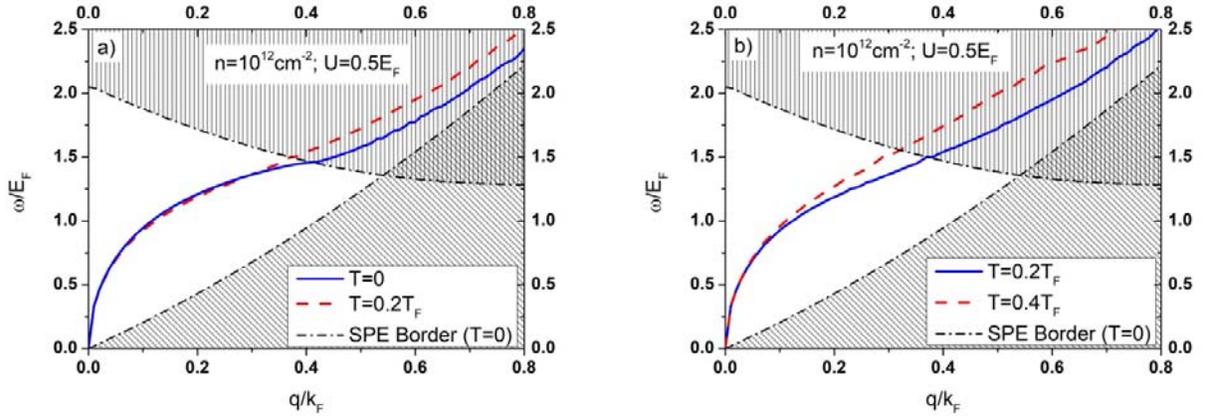

**Fig. 1.** Plasmon frequencies in a BBLG with $n = 10^{12} cm^{-2}$, $U = 0.5E_F$ for $T = 0$ and $0.2T_F$ (a), and $T = 0.2$ and $0.4T_F$ (b). Dashed-dotted lines show the boundaries of the single-particle-excitation (SPE) continuum.

To see in more detailed the impact of temperature, we plot in Fig. 2 the plasmon frequency in a BBLG, with $n = 10^{12} cm^{-2}$, $U = E_F$ (a) and $U = 0$ (b), as a function of temperature varying from zero to $0.8T_F$ for several values of wave vector. We observe that the plasmon frequency decreases slightly with temperature for $T < T_C$ and then increases remarkably at higher $T$. The critical temperature $T_C$ increases (decreases) notably with potential bias $U$ (wave number $q$). We also find that, as in some layered structures consisting of graphene [28, 34, 35], the temperature has larger impact on the plasmon energy at larger wave vectors.

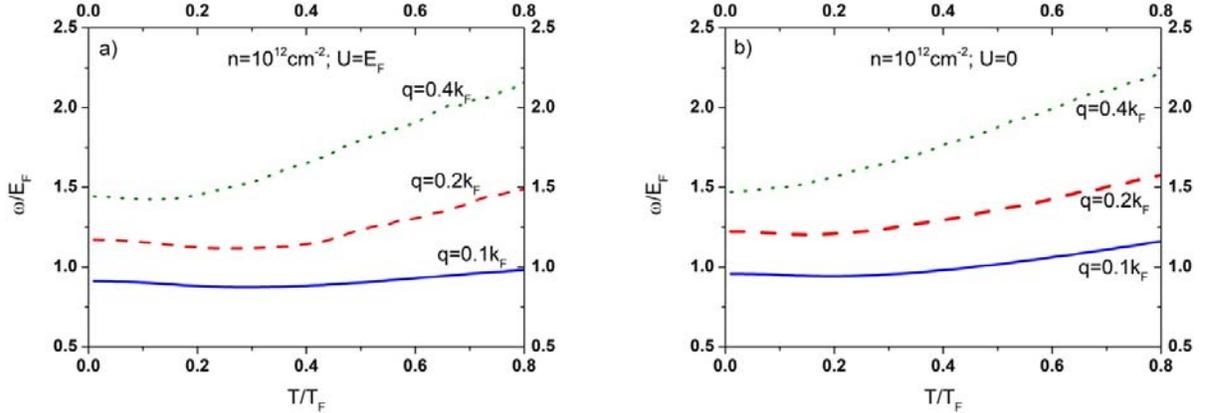

**Fig. 2.** Plasmon frequency in a BBLG, with $n = 10^{12} cm^{-2}$, $U = E_F$ (a) and $U = 0$ (b), as a function of temperature for several values of wave vector.

We now study the combined effect of finite temperatures and potential bias on collective excitations by showing in Fig. 3 the plasmon dispersions in a BBLG under potential bias $U = 0.5E_F$ and $U = 1.5E_F$ for two cases $T = 0.2T_F$ (a) and $T = 0$ (b). It is seen from the figures that, in both cases, the increase in energy gap decreases notably plasmon frequency as in gapped graphene structures [35-40]. The bias effect is somewhat stronger for larger temperature especially in the large momentum region. Note that the increase of the effect of dielectric inhomogeneity, on plasmon modes, with temperature has been found in Ref. 30. At $U = 0.5E_F$ ($U = 1.5E_F$) the plasmon mode enters into the interband SPE continuum and becomes more damped (the plasmon mode touches the intraband SPE edge and disappears) at smaller wave vector compared to the zero-temperature case.

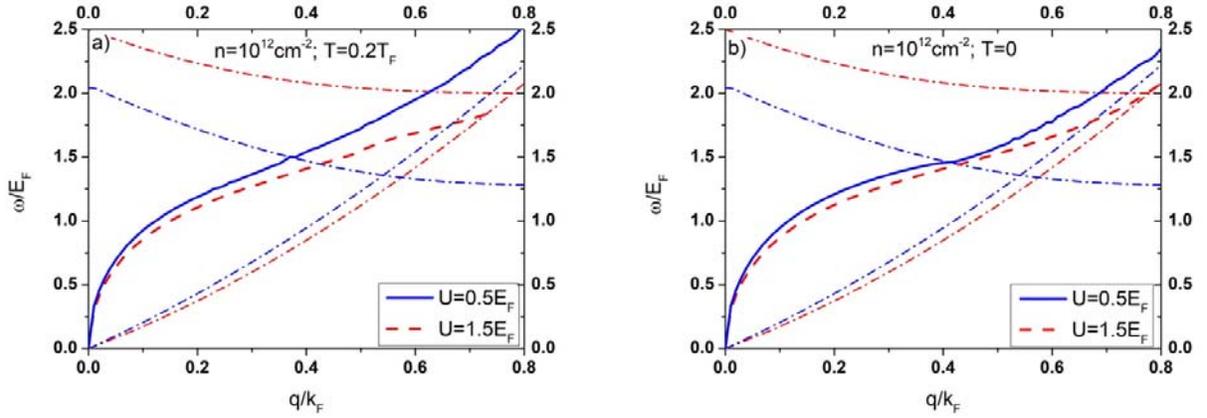

**Fig. 3.** Plasmon dispersions in a BBLG with $n=10^{12} cm^{-2}$, $U=0.5E_F$ and $U=1.5E_F$ for $T=0.2T_F$ (a) and $T=0$ (b). Dashed-dotted lines show the boundaries of SPE continuum.

Finally, to understand better how doping level affects collective excitations in zero- and finite temperature BBLG systems we plot in Fig. 4 plasmon dispersions for $U=1.5E_F$, $n=10^{12} cm^{-2}$ and $n=10^{13} cm^{-2}$ at $T=0.2T_F$ (a) and $T=0$ (b). The figures indicate that the increase in carrier density in BBLG decreases dramatically plasmon frequency (in unit of Fermi energy). At low temperature and large bias the finite-temperature plasmon mode touches the intraband SPE edge and disappears at smaller wave vectors in comparison with the zero-temperature case. Hence the collective excitation mode is less observable in experiments at finite temperatures as expected.

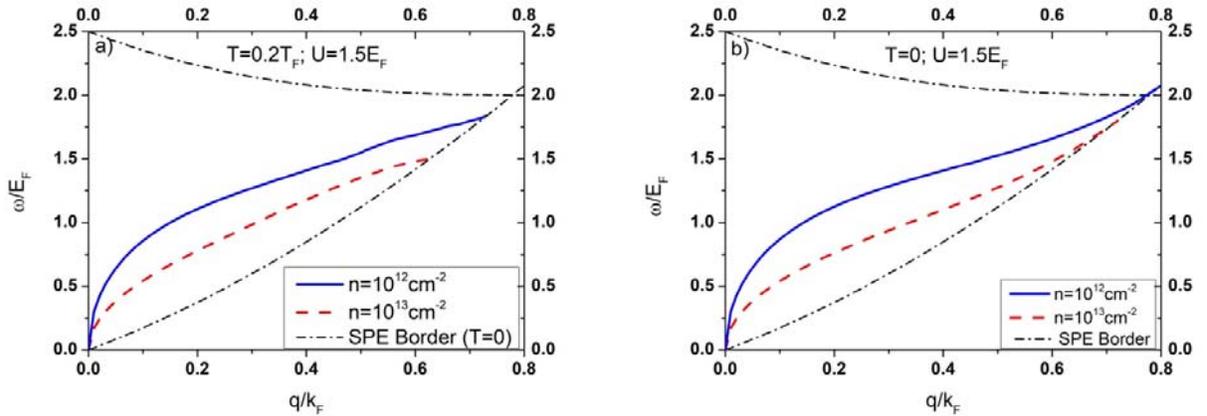

**Fig. 4.** Plasmon dispersions in a BBLG with $U=1.5E_F$, $n=10^{12} cm^{-2}$ and $n=10^{13} cm^{-2}$ for $T=0.2T_F$ (a) and $T=0$ (b). Dashed-dotted lines show the boundaries of SPE continuum.

## 4. Conclusion

In summary, collective excitations in the BBLG at finite temperature $T$ have been calculated for the first time. The obtained results indicate that an weakly damped collective mode with dispersion curve outside the SPE continuum exists in the system as in the zero-temperature case. We find that for small (large) wave vectors, the temperature modifies slightly (increases significantly) the plasmon frequency. The increase in potential bias decreases remarkably plasmon frequency and the bias effect is somewhat stronger for larger temperature especially in the large momentum region. At finite-temperature and small (large) bias the plasmon mode enters into the interband SPE continuum and becomes more damped (the plasmon mode touches the intraband SPE edge and disappears) at smaller wave vector compared to the zero-temperature case. This effect makes the collective excitation mode observable in experiments at large bias becomes less observable at finite temperatures.